\documentclass[prd,twocolumn,preprintnumbers,superscriptaddress,floatfix,nofootinbib,amsmath,amssymb]{revtex4} 
\usepackage{graphicx,color,dcolumn,booktabs,bm,mathrsfs}
\usepackage{longtable,lscape}
\usepackage{txfonts}
\usepackage{overpic}
\usepackage{amssymb}
\usepackage{indentfirst}
\usepackage{feynmf}   
\usepackage{slashed}  
\usepackage{cases}
\usepackage{color}
\usepackage{multirow}
\usepackage{epstopdf}
\usepackage{CJK}

\usepackage[colorlinks,
            citecolor=blue,
            anchorcolor=red,
            menucolor=red,
            linkcolor=red,
            filecolor=red,
            runcolor=red,
            urlcolor=blue,
            frenchlinks=red]{hyperref}
\usepackage{subfigure}

\newcommand{\mev}{\textrm{ MeV}}
\newcommand{\proc}{$B^+ \to D^{*-}D^+K^+$ }
\newcommand{\dk}{$D^+K^+$ }

\begin{document}

\title{Possible signal of an exotic $I=1$, $J=2$ state in the $B \to D^{*-}D^+K^+$ reaction}

\author{Wen-Tao Lyu}
\affiliation{Department of Physics, Guangxi Normal University, Guilin 541004, China}
\affiliation{Guangxi Key Laboratory of Nuclear Physics and Technology, Guangxi Normal University, Guilin 541004, China}
\affiliation{School of Physics, Zhengzhou University, Zhengzhou 450001, China}

\author{Man-Yu Duan}\email{duanmy@seu.edu.cn}
\affiliation{School of Physics, Southeast University, Nanjing 210094, China}
\affiliation{School of Physics, Zhengzhou University, Zhengzhou 450001, China}
\affiliation{Departamento de Física Teórica and IFIC, Centro Mixto Universidad de Valencia-CSIC Institutos de Investigación de Paterna, 46071 Valencia, Spain}

\author{Chu-Wen Xiao}\email{xiaochw@gxnu.edu.cn}
\affiliation{Department of Physics, Guangxi Normal University, Guilin 541004, China}
\affiliation{Guangxi Key Laboratory of Nuclear Physics and Technology, Guangxi Normal University, Guilin 541004, China}
\affiliation{School of Physics, Central South University, Changsha 410083, China}

\author{En Wang}\email{wangen@zzu.edu.cn}
\affiliation{School of Physics, Zhengzhou University, Zhengzhou 450001, China}
\affiliation{Guangxi Key Laboratory of Nuclear Physics and Technology, Guangxi Normal University, Guilin 541004, China}

\author{Eulogio Oset}\email{oset@ific.uv.es}
\affiliation{Department of Physics, Guangxi Normal University, Guilin 541004, China}
\affiliation{Departamento de Física Teórica and IFIC, Centro Mixto Universidad de Valencia-CSIC Institutos de Investigación de Paterna, 46071 Valencia, Spain}
\affiliation{School of Physics, Zhengzhou University, Zhengzhou 450001, China}\vspace{0.5cm}

\begin{abstract}

We study the \proc reaction, showing that a peak in the \dk mass distribution around $2834 \mev$ reported by LHCb could be associated with a theoretical exotic state with that mass, a width of $19 \mev$ and $J^P=2^+$, stemming from the interaction of the  $D^{*+}K^{*+}$ and $D^{*+}_s \rho^+$ channels, which is a partner of the $0^+$ $T_{c\bar{s}}(2900)$. We show that the data is compatible with this assumption, but also see that the mass distribution itself cannot discriminate between the spins $J=0$, $1$, $2$ of the state. Then we evaluate the momenta of the angular mass distribution and show that they are very different for each of the spin assumptions, and that the momenta coming from interference terms have larger strength at the resonant energy than the peaks seen in the angular integrated mass distribution. We make a call for the experimental determination of these magnitudes, which has already been used by the LHCb in related decay reactions.

\end{abstract}

\date{\today}

\maketitle


\section{INTRODUCTION}
\label{sec:INTRODUCTION}

The study of the vector vector interaction has attracted much attention after the pioneering works of Refs.~\cite{Molina:2008jw,Geng:2008gx} in the light quark sector \cite{Molina:2009ct,Molina:2010tx,Molina:2009eb,Wang:2011tm,Aceti:2014kja,Gulmez:2016scm,Du:2018gyn,Geng:2016pmf,Wang:2019niy,Molina:2019rai,Molina:2020hde,He:2020btl,Montana:2022inz,Duan:2023lcj,Shen:2024jfr}. In the light quark sector the vector-vector interaction gives rise to resonances as the $f_2(1270)$, $f_0(1370)$, $f'_2(1525)$, $f_0(1710)$, $a_0(1710)$, $K_2^*(1430)$. A recent review of applications of these results to different reaction can be found in Ref.~\cite{Shen:2024jfr}. The $a_0(1710)$ was a prediction of the work of Ref.~\cite{Geng:2008gx}, which was found recently in Refs.~\cite{BaBar:2021fkz,BESIII:2021anf} (see related works in Refs.~\cite{Dai:2021owu,Oset:2023hyt,Wang:2022pin,Zhu:2022wzk,Bayar:2022wbx,Dai:2022qwh,Zhu:2022guw,Abreu:2023xvw,Wang:2023aza,Achasov:2023izs,Wang:2023lia,Dai:2022htx,Ding:2023eps,Ding:2024lqk,Peng:2024ive}). In Ref.~\cite{Molina:2009ct}, the formalism of the interaction based upon the local hidden gauge \cite{Bando:1984ej,Bando:1987br,Harada:2003jx,Meissner:1987ge,Nagahiro:2008cv} with unitary in coupled channels, was extended to the charm sector and several states were also generated as molecular states of $D^*\bar{D}^*$ or $D_s^*\bar{D}_s^*$ with different isospin and spin $J=0$, $1$, $2$~\cite{Duan:2022upr,Duan:2021pll,Liu:2020ajv,Zhang:2020rqr,Dai:2018nmw,Wang:2018djr,Wang:2017mrt}.

One of the interesting works, for its posterior repercussion, was the one of Ref.~\cite{Molina:2010tx}. Indeed, in this work among other states, it was found that $D^*\bar{K}^*$ in isospin $I=0$ should lead to a bound state in spin parity $J^P=0^+$, $1^+$, $2^+$. The $J=0^+$ state was little bound and the $1^+$, $2^+$, particularly the $2^+$ state, more bound, all of them with small width of the order $20-50 \mev$. The state with these characteristics with $0^+$ was found by the LHCb collaboration \cite{LHCb:2020bls,LHCb:2020pxc}, by looking at the $D^-K^+$ invariant mass distribution, and named $X_0(2900)$ (now called $T_{\bar{c}\bar{s}0}(2870)$). The state with $D^-K^+$ is clearly exotic since it corresponds to a $\bar{c}\bar{s}du$ quark configuration which cannot be accommodated by a standard $q\bar{q}$ state. A follow up of this finding, refining the input of Ref.~\cite{Molina:2010tx} to the light of the experimental data, was done in Ref.~\cite{Molina:2020hde}, where more precise predictions for mass and width of the unobserved $1^+$, $2^+$ states were done, suggesting the reactions where they could be found. This issue has also had a warm reception in the recent literature and work along these lines can be seen in the references of Ref.~\cite{Ding:2024dif}.

Another prediction of an exotic state made in Ref.~\cite{Molina:2010tx} was a state generated by the interaction of the $D^*K^*$ and $D_s^* \rho$ coupled channels, with $I=1$, which in quark language would correspond to $c\bar{s}u\bar{d}$, once again an exotic state. The state was obtained as a cusp between the $D_s^* \rho$ and $D^*K^*$ thresholds. A state of these characteristics was observed by the LHCb collaboration in Ref.~\cite{LHCb:2022lzp} in the $D_s^+\pi^-$, $D_s^+\pi^+$ mass distributions of the $B^0 \to \bar{D}^0 D_s^+ \pi^-$ and $B^+ \to D^- D_s^+ \pi^+$ reactions. The state is now called $T_{c\bar{s}}(2900)$ with $I=1$ and $J^P=0^+$. An update of Ref.~\cite{Molina:2010tx} of this state to the light of the LHCb data is done in Ref.~\cite{Molina:2022jcd}, where once again the $T_{c\bar{s}}(2900)$ appears as a threshold effect of the $D_s^* \rho$ channel. A follow up of works on the issue can be seen in Refs.~\cite{Lyu:2024wxa,Lyu:2023aqn,Lyu:2023ppb,Duan:2023qsg}. Yet, the striking thing from the results of Ref.~\cite{Molina:2022jcd} was that the state with $1^+$ was also found as a cusp, but the state with $2^+$ was found bound by about $86 \mev$. The stronger weight of the interaction in the $2^+$ channel of the vector vector interaction is a constant in the different problems studied, starting from the $f_2(1270)$ generated in the $\rho \rho$ interaction, which is more bound than the $f_0(1370)$ \cite{Molina:2008jw}.

The results of Ref.~\cite{Molina:2022jcd} concerning this $2^+$ state are more reassuring when one compares them with similar, independent studies in Ref.~\cite{Duan:2023lcj}. There a mass for this state running from $2780-2866 \mev$, corresponding to a binding of $114-140 \mev$ was found. The width for this state was found of $19 \mev$ in Ref.~\cite{Molina:2022jcd}, while it was not evaluated in Ref.~\cite{Duan:2023lcj}. Since the results of Ref.~\cite{Molina:2022jcd} were constrained by data from Refs.~\cite{LHCb:2022lzp,LHCb:2020bls,LHCb:2020pxc}, we can consider them as a more precise determination, and together with the width of $19 \mev$, a genuine prediction of a state to be observed. The success of the observation of previous predicted states, as described above gives us confidence about finding this state in the future.

With this conviction we want to call the attention here to some experimental information that provides hints of this state. The information comes from the LHCb experiment \cite{LHCb:2024vfz}, where the authors studied the $B^+ \to D^{*-} D^+ K^+$ reaction among others. In Fig.~1(c) of Ref.~\cite{LHCb:2024vfz} there is a prominent peak  in the $D^+ K^+$ mass distribution, relatively narrow, around $2835 \mev$. While the errors are relatively large and one strong fluctuation could not be excluded, the structure is clearly visible and the coincidence of this energy with the prediction of Ref.~\cite{Molina:2022jcd} makes it more intriguing. If one adds to that a peak (with poor statistics) also visible in the \dk invariant mass distribution of the $B^+ \to D^+ D^- K^+$ reaction of Ref.~\cite{LHCb:2020pxc} in Figs. 5(d) and 6(d), the case gets further support and we take the information of Ref.~\cite{LHCb:2024vfz} to test the theoretical predictions. We see that the predictions of the mass and width fit well with the data, but we also see that the data does not discriminate between a $J=0$, $1$, $2$ state. In view of this, we propose to go one step forward in the experiment and construct the moments $\frac{d\Gamma_l}{dM_{\mathrm{inv}}} = \int d\Omega \frac{d\Gamma}{dM_{\mathrm{inv}} d\Omega} Y_{l0}$, which have occasionally been evaluated in experimental analyses. Indeed, in the LHCb experiment of Ref.~\cite{LHCb:2020pxc} on the $B^+ \to D^+ D^- K^+$ reaction, the moments of different values of $l$ for the different pairs are evaluated. While the errors are big and there are obvious fluctuations, the different moments have some clear structure. Actually, in Ref.~\cite{Bayar:2022wbx} the $B^+ \to D^+ D^- K^+$ reaction was suggested to observe the $J^P=2^+$ partner of the $X_0(2900)$ by looking at the moments of the \dk mass distribution. In this case there was interference of the $X_1(2900)$ with the $J^P=2^+$ state and a particular pattern was obtained for $\frac{d\Gamma_l}{dM_{\mathrm{inv}}}$ of $l=3$, which agreed remarkably well with the experimental analysis. Yet, the case was not strong enough to make a claim, given the fact that the LHCb analysis without the $J^P=2^+$ state also gave a good reproduction of the momenta \cite{gershon}. In Ref.~\cite{Wang:2021naf}, the idea was also used when studying the $D^+_s \to K^+ K^- \pi^+$ reaction.

In the present case we have a state with $2^+$, far away from the other $0^+$, $1^+$ states, and we expect only interference with a tree level contribution. We make calculations of different moments for \dk up to $l=4$ and see that in that case one could discriminate between $J=0$, $1$, $2$. The fact that we have now \proc instead of $B^+ \to D^+ D^- K^+$ as in Ref.~\cite{LHCb:2020pxc} the pattern of resonance of $D^{*-}D^+$ or  $D^{*-}K^+$ resonance, and their possible replicas in the \dk mass distribution are very different to those of $D^-D^+$, $D^-K^+$, hence, the information is complementary, and given the bigger masses of $D^{*-}D^+$ and  $D^{*-}K^+$ possible resonance states compared to those of $D^-D^+$, $D^-K^+$, the pattern of \dk momenta in the \proc reaction could be simpler than that in $B^+ \to D^+ D^- K^+$. Given the compelling theoretical support for the $2^+$ state, the work done here should be an incentive for this experimental analysis to be performed.


  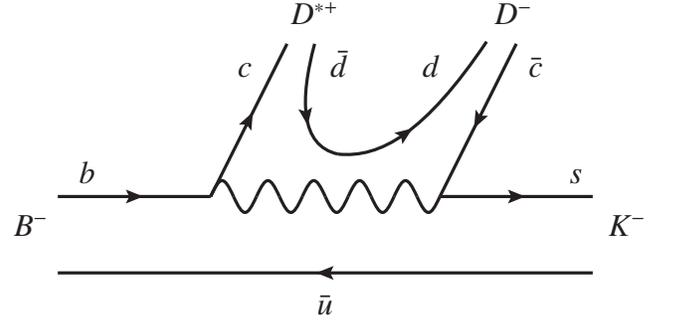
\begin{figure}[htb]
  \centering
 \includegraphics[scale=0.8]{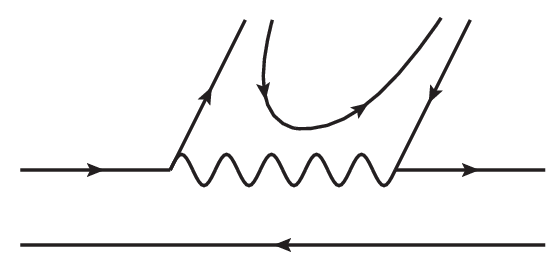}\put(-225,20){\large{$B^-$}}\put(0,20){\large{$K^-$}}\put(-110,-10){\large{$\bar{u}$}}\put(-200,40){\large{$b$}}\put(-15,40){\large{$s$}}\put(-140,80){\large{$c$}}\put(-105,80){\large{$\bar{d}$}}\put(-70,80){\large{$d$}}\put(-30,80){\large{$\bar{c}$}}\put(-120,100){\large{$D^{*+}$}}\put(-43,100){\large{$D^{-}$}}
  \caption{Internal emission mechanism for $B^- \to D^{*+}D^- K^-$.}
  \label{Fig:mech}
  \end{figure}
  
 \section{FORMALISM}
\label{sec:FORMALIISM}

The process of \proc decay can proceed via internal emission \cite{Chau:1982da} as shown in Fig.~\ref{Fig:mech}, for the charge conjugate channel. As we can see, it is possible to produce $D^{*+}D^- K^-$ at the tree level. Recall that originally we have $B^+ \to D^{*-} D^{*+} K^{*+}$ production. Since angular momentum is conserved in the weak decay, and we have $D^{*+}$ with $J^P=1^-$, $B^-$ with $0^-$, and $D^{*-} K^{*-}$ in $S$-wave, respectively, where we need a $P$-wave mechanism to govern the reaction, contracting $\epsilon^{\mu}(D^{*+})$ with a vector.

Coming back to \proc, since we want \dk coming from $D^{*+} K^{*+}$ in $S$-wave, producing the $2^+$ resonance, which later decays in $D$-wave in \dk, then the $\epsilon_{\mu}$ polarization of the $D^{*-}$ has to be contracted with $P_B^{\mu}$.

Following Ref.~\cite{Bayar:2022wbx} in the study of the analogous $B^+ \to D^- D^+ K^+$ reaction, we write the amplitude for the transition \proc,
\begin{eqnarray}\label{t}
t=\epsilon_{\mu}(D^{*-})P_B^{\mu}(aY_{00}+bY_{20}+cY_{10}),
\end{eqnarray}
assuming that we have first produced $D^{*-} D^{*+} K^{*+}$ and then $D^{*+} K^{*+} \to D^+ K^+$ in $S$-wave ($a$ term in Eq.~\eqref{t}), $D$-wave ($b$ term in Eq.~\eqref{t}) and $P$-wave ($c$ term in Eq.~\eqref{t})

The mass and angular distribution is given by
\begin{eqnarray}\label{8-2}
\frac{d\Gamma}{dM_{\mathrm{inv}}(D^+K^+)d\tilde{\Omega}}=\frac{1}{(2\pi)^4}\frac{1}{8M_B^2}p_{D^{*-}} \tilde{k} \sum|t|^2,
\end{eqnarray}
where $\tilde{\Omega}$ is the solid angle of \dk in their rest frame, and
\begin{eqnarray}
p_{D^{*-}}&=&\frac{\lambda^{1/2}\left(M_B^2, m_{D^{*-}}^2, M^2_{\mathrm{inv}}(D^+K^+)\right)}{2M_B}, \nonumber\\
\tilde{k}&=&\frac{\lambda^{1/2}\left(M^2_{\mathrm{inv}}(D^+K^+), m_D^2, m_K^2\right)}{2M_{\mathrm{inv}}(D^+K^+)},
\end{eqnarray}
We easily find for the sum over the $D^{*-}$ polarization in $|t|^2$,
\begin{eqnarray}
 \sum|t|^2 &=& \left( \frac{ M_{B^+}}{M_{D^{*-}}} \right)^2 \vec{p}^2_{D^{*-}} \Big( |a|^2 Y_{00}^2 + |b|^2 Y_{20}^2 + |c|^2 Y_{10}^2 \nonumber\\
 &+& 2Re(ab^*) Y_{00} Y_{20} + 2 Re(ac^*) Y_{00} Y_{10}  \nonumber\\
 &+& 2Re(bc^*) Y_{20}Y_{10}\Big),
\end{eqnarray}
We then define the moments ,
\begin{eqnarray}
\frac{d\Gamma_l}{dM_{\mathrm{inv}}} = \int d \tilde{\Omega} \frac{d\Gamma}{dM_{\mathrm{inv}} d \tilde{\Omega}} Y_{l0},
\end{eqnarray}
from Eq.~\eqref{8-2} it is easy to find the relations,

\begin{eqnarray}\label{gammal}
\dfrac{d~\Gamma_{0}}{dM_{\rm inv} }&=&FAC \left[ \vert a \vert^{2} +\vert b \vert^{2} +\vert c \vert^{2} \right]  , \nonumber\\
\dfrac{d~\Gamma_{1}}{dM_{\rm inv} }&=&FAC \left[ 2~ Re(ac^{*}) + \frac{2}{\sqrt{5}} ~2~ Re(bc^{*}) \right]   , \nonumber\\
\dfrac{d~\Gamma_{2}}{dM_{\rm inv} }&=&FAC \left[ \frac{2}{7}\sqrt{5} ~\vert b \vert^{2} + \frac{2}{5}\sqrt{5} ~\vert c \vert^{2} +2~ Re(ab^{*})\right]   , \nonumber\\
\dfrac{d~\Gamma_{3}}{dM_{\rm inv} }&=&FAC  \sqrt{\frac{15}{7}} ~\frac{3}{5} ~2~ Re(bc^{*})  , \nonumber\\
\dfrac{d~\Gamma_{4}}{dM_{\rm inv} }&=&FAC ~\frac{6}{7} \vert b \vert^{2}  , 
\end{eqnarray}
where
\begin{equation}
FAC =\frac{1}{\sqrt{4 \pi}} \frac{1}{(2 \pi)^4} \frac{1}{8 M^{2}_{B^+}} ~ \vec{p}^2_{D^{*-}}~p_{D^{*-}}~ \tilde{k} ~\left(\frac{M_{B^+}}{M_{D^{*-}}}\right)^2 . 
\end{equation}
Hence,
\begin{equation}
\dfrac{d~\Gamma}{dM_{\rm inv} }=\sqrt{4\pi} \dfrac{d~\Gamma_{0}}{dM_{\rm inv} }.
\end{equation}
Equation~\eqref{gammal} agree with Refs.~\cite{Bayar:2022wbx,LHCb:2016lxy}.


\section{RESULTS AND DISCUSSIONS}
\label{sec:RESULTS}

We show the results in three different scenarios: (I), assuming that we have an $S$-wave resonance, $a \neq 0$, $b=c=0$; (II), assuming that we have a $P$-wave resonance, $a \neq 0$, $b=0$, $c \neq 0$ (Note that we always keep a tree level $S$-wave amplitude with $a \neq 0$); (III), assuming that we have a $D$-wave resonance (our preferred choice from present theoretical calculations),  $a \neq 0$, $b \neq 0$, $c = 0$.

\begin{itemize}

\item Case I:
We have now from Eq.~\eqref{t}
\begin{equation}\label{case1}
aY_{00}=\left(a_0 + a'_0 \frac{M_B^2}{M^2_{\rm inv}(D^+K^+)-M_R^2+i M_R \Gamma_R}\right) Y_{00},
\end{equation}
where $a_0$ stands for the tree level contribution and the term $a'_0$ gives the weight of the resonance. The factor $M_B^2$ in Eq.~\eqref{case1} is put to have $a_0$ and $a'_0$ with the same dimensions. We assume in all cases that the resonance has the mass and width found in Ref.~\cite{Molina:2022jcd} and hinted by the experiment \cite{LHCb:2024vfz}, $M_R=2834 \mev$, $\Gamma_R=19 \mev$. We fit then $a_0$ and $a'_0$ to the experiment.

\item Case II:
From Eq.~\eqref{t} we have now
\begin{eqnarray}\label{case2}
&&aY_{00}+cY_{10} \nonumber\\
&&=a_1Y_{00} + c' \frac{M_B \tilde{k}}{M^2_{\rm inv}(D^+K^+)-M_R^2+i M_R \Gamma_R}Y_{10},
\end{eqnarray}
where we have put the factor $\tilde{k}$ suited to a $P$-wave amplitude. We fit again $a_1$ and $c'$ to the data on $d \Gamma / d M_{\rm inv}$ and then calculate the different moments.

\item Case III:
From Eq.~\eqref{t} we have now
\begin{eqnarray}\label{case3}
&&aY_{00}+bY_{20} \nonumber\\
&&=a_2Y_{00} + b' \frac{\tilde{k}^2}{M^2_{\rm inv}(D^+K^+)-M_R^2+i M_R \Gamma_R}Y_{20},
\end{eqnarray}
where, again, we implement explicitly the factor $\tilde{k}^2$ suited to the $D$-wave resonance. We fit $a_2$ and $b'$ to experiment and then calculate the momenta.

\end{itemize}

Since in cases II and III, the $S$-wave from the background tree level does not interfere with the $P$ or $D$-wave in the angle integrated mass distribution, then $a_1$ and $a_2$ are the same. In the case I, the $a_0$ and $a'_0$ terms will interfere. In the case II, we expect to here an interference term from $Re(ac^*)$ in $\frac{d~\Gamma_{1}}{dM_{\rm inv} }$ and $\frac{d~\Gamma_{3}}{dM_{\rm inv} }=\frac{d~\Gamma_{4}}{dM_{\rm inv} }= 0$. In the case III, we expect $\frac{d~\Gamma_{1}}{dM_{\rm inv} }= 0$, and $\frac{d~\Gamma_{3}}{dM_{\rm inv} }= 0$, while here should be an interference term from $2Re(ab^*)$ in $\frac{d~\Gamma_{2}}{dM_{\rm inv} }$.

  \begin{figure}[htb]
  \centering
 \includegraphics[scale=0.7]{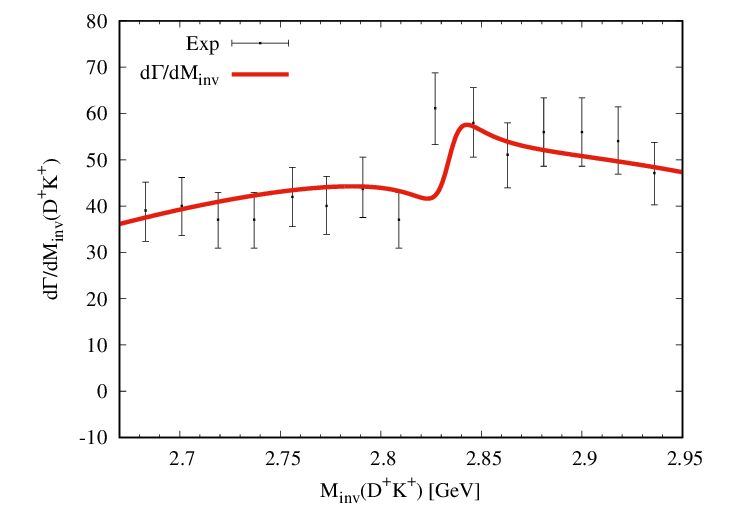}
  \caption{Fit to $\dfrac{d~\Gamma}{dM_{\rm inv} }(D^+K^+)$ in case I:\, $b = c = 0$.}
  \label{Fig:case1}
  \end{figure}
  
In Fig.~\ref{Fig:case1} we show the results for the fit in case I. In all cases we have found that a good background tree level is obtained with an empirical amplitude changing
\begin{equation}
a_i \to \tilde{a}_i \frac{\tilde{k}}{M_B},
\end{equation}
with this change we see the results in Fig.~\ref{Fig:case1}. We see that a fair reproduction of the data is found with the parameters,
\begin{eqnarray}
\tilde{a}_0&=& 13.83 \mev^{-1} \, , \nonumber\\
a'_0&=& 5.48\times10^{-4} \mev^{-1} \, .
\end{eqnarray}
In this case all the moments $\frac{d~\Gamma_{l}}{dM_{\rm inv} }$ for $l \neq 0$ are zero. One could not rule out such a state based on $\frac{d~\Gamma}{dM_{\rm inv} }$, but the moments $\frac{d~\Gamma_{l}}{dM_{\rm inv} }$ would be discriminating.

  \begin{figure}[htb]
  \centering
 \includegraphics[scale=0.7]{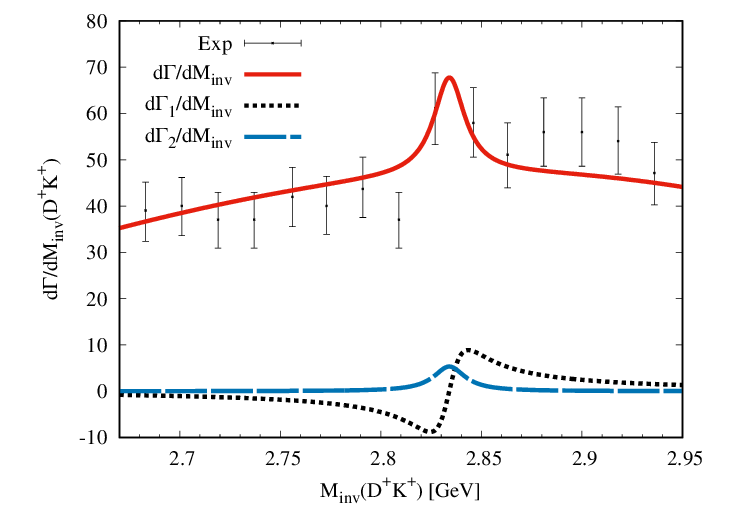}
  \caption{Fit to $\dfrac{d~\Gamma}{dM_{\rm inv} }(D^+K^+)$ for case II:\, $b = 0$, $c \neq 0$.}
  \label{Fig:case2}
  \end{figure}

In Fig.~\ref{Fig:case2} we show the results for case II, where we assume that the state could correspond to $J=1$, with the resonance decaying in $L=1$ to \dk. The results of the fit to $\frac{d~\Gamma}{dM_{\rm inv} }$ are also acceptable with the fit parameters,
\begin{eqnarray}
\tilde{a}_1&=& 13.49 \mev^{-1} \, , \nonumber\\
c'&=& 1.75\times10^{-2} \mev^{-1} \, .
\end{eqnarray}
Now we see that the moments play a role. $\frac{d~\Gamma_1}{dM_{\rm inv} }$ now has an interference pattern, going from positive to negative, and $\frac{d~\Gamma_2}{dM_{\rm inv} }$ always positive. Interestingly, since the interference term in $\frac{d~\Gamma_1}{dM_{\rm inv} }$ is linear in $c'$, while $\frac{d~\Gamma_2}{dM_{\rm inv} }$ is quadratic in $c'$, the strength of $\frac{d~\Gamma_1}{dM_{\rm inv} }$ is bigger than that of $\frac{d~\Gamma_2}{dM_{\rm inv} }$, or $\frac{d~\Gamma}{dM_{\rm inv} }$ (dividing by $\sqrt{4 \pi}$). In other words, the use of the momentum magnitude $\frac{d~\Gamma_1}{dM_{\rm inv} }$ has stressed the signal of the resonance versus the one obtained from $\frac{d~\Gamma}{dM_{\rm inv} }$.

  \begin{figure}[htb]
  \centering
 \includegraphics[scale=0.7]{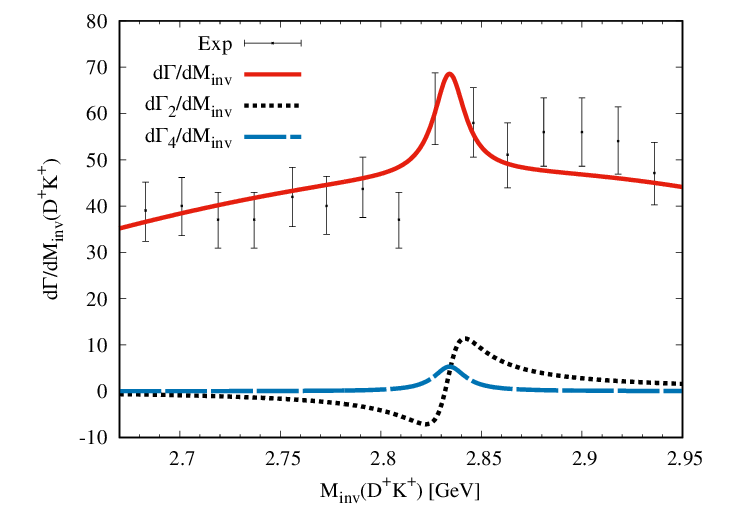}
  \caption{Fit to $\dfrac{d~\Gamma}{dM_{\rm inv} }(D^+K^+)$ for case III:\, $c = 0$, $b \neq 0$.}
  \label{Fig:case3}
  \end{figure}
  
In Fig.~\ref{Fig:case3} we show the results for case III with the resonance $2^+$ decaying to \dk in $D$-wave. The results are similar, we can get a good fit to $\frac{d~\Gamma}{dM_{\rm inv} }$ with the parameters,
\begin{eqnarray}
\tilde{a}_2&=& 13.48 \mev^{-1} \, , \nonumber\\
b'&=& 1.38 \times10^{-1} \mev^{-1} \, .
\end{eqnarray}
But now it is $\frac{d~\Gamma_2}{dM_{\rm inv} }$ the magnitude, linear in $b'$, that shows the interference and $\frac{d~\Gamma_4}{dM_{\rm inv} }$ is quadratic in $b'$. Once again the interference magnitude has a bigger strength than $\frac{d~\Gamma_4}{dM_{\rm inv} }$, or $\frac{d~\Gamma}{dM_{\rm inv} }\frac{1}{\sqrt{4\pi}}$. We should note that we have taken $\tilde{a}_1$, $\tilde{a}_2$, $c'$, $b'$ of the same sign. Should we reverse the relative sign of $c'$, $b'$ versus $\tilde{a}_1$, $\tilde{a}_2$ the pattern of the interference magnitude is the same except that the sign changes.

We can see that the pattern of the interference moments $\frac{d~\Gamma_l}{dM_{\rm inv} }$ changes from assuming a $J=1$ or a $J=2$ state, and for $J=0$ there are no moments except $\frac{d~\Gamma_0}{dM_{\rm inv} }$. In other words, a careful study of the different moments associated to the angular dependent mass distribution should determine the spin of the resonance. We think that the results found here should be a sufficient motivation to carry on the determination of these magnitudes, as it was done for the $B^+ \to D^- D^+ K^+$ reaction in Ref.~\cite{LHCb:2020pxc}.


\section{Summary}

We have studied the \proc reaction, looking for the contribution of a particular channel, $B^+ \to D^{*-} D^{*+} K^{*+}$, with decay of $D^{*+}K^{*+}$ to \dk. The idea of the reaction comes from the fact that in the study of the $D^{*+}K^{*+}$ interaction with its coupled channel $D^*_s \rho$, apart from the signal for the $J^P=0^+$ $T_{c\bar{s}}(2900)$ state, appearing as a threshold effect in the $D^{*+}K^{*+}$ and $D^*_s \rho$ channels, a clear bound state is obtained in the $2^+$ sector at $2834 \mev$, bound by about $80 \mev$. This state is clearly exotic and decays to \dk, where it should be observed. We call the attention to the fact that in the LHCb experimental data for the \proc reaction, some signal is seen precisely in the \dk invariant mass distributions at that energy, and hints are also observed at the same energy in the $B^+ \to D^- D^+ K^+$ reaction also measured by the LHCb collaboration. We carry out fits to the data on the \dk mass distribution assuming a mass of $2834 \mev$ and width of $19 \mev$ for the resonance, as obtained by the theory, and we find a good agreement with the data. Yet, we also show that the invariant mass distribution by itself cannot discriminate between having spin $0$,$1$,$2$ for the resonance, and then we propose to use the momenta associated to the angular dependent mass distribution. These magnitudes have been determined experimentally for the  $B^+ \to D^- D^+ K^+$ reaction, but not for the \proc one. We then calculate the different momenta up to $l=4$ and show that they are drastically different for each of the spin assumptions. We also show that the momenta that involve interference of amplitudes, which is linear with the coefficient carried by the resonance term in the transition amplitude, have a strength bigger than the signal in the angular integrated mass distribution, where this coefficient appears quadratic. 
      
In summary we show how the determination of the moments of the \dk  mass distribution, can lead to learn whether there is indeed a resonance associated with the peak observed in the experiment, and which is the spin of the exotic resonance.  
      
\label{sec:Summary}

\section*{Acknowledgement}
This work is supported by the National Key Research and Development Program of China (No. 2024YFE0105200). This work is supported by the Natural Science Foundation of Henan under Grant No. 232300421140 and No. 222300420554, the National Natural Science Foundation of China under Grant No. 12475086 and No. 12192263. This work is also supported by the Spanish Ministerio de Economia y Competitividad (MINECO) and European FEDER funds under Contracts No. FIS2017-84038-C2-1-P B, PID2020-112777GB-I00, and by Generalitat Valenciana under contract PROMETEO/2020/023. This project has received funding from the European Union Horizon 2020 research and innovation programme under the program H2020-INFRAIA-2018-1, grant agreement No. 824093 of the STRONG-2020 project, and partly by the Natural Science Foundation of Changsha under Grant No. kq2208257, the Natural Science Foundation of Hunan Province under Grant No. 2023JJ30647, the Natural Science Foundation of Guangxi Province under Grant No. 2023JJA110076, and the National Natural Science Foundation of China under Grants No. 12365019 (CWX).


\end{document}